**The Journal of Phytopharmacology**
(Pharmacognosy and phytomedicine Research)

**Research Article**

**HKK Rajapaksha**
Department of Basic Sciences, Faculty of Allied Health Sciences, General Sir John Kotelawala Defence University, Werahera, Sri Lanka

**MN Fernando**
Department of Pharmacy, Faculty of Allied Health Sciences, General Sir John Kotelawala Defence University, Sri Lanka

**NRM Nelumdeniya**
Department of Pharmacy, Faculty of Allied Health Sciences, General Sir John Kotelawala Defence University, Sri Lanka

**AWMKK Bandara**
Department of Basic Sciences, Faculty of Allied Health Sciences, General Sir John Kotelawala Defence University, Sri Lanka

**ARN Silva**
Department of Basic Sciences, Faculty of Allied Health Sciences, General Sir John Kotelawala Defence University, Weraera, Sri Lanka

# Evaluation of *In vitro* anti-inflammatory activity and *In-silico* pharmacokinetics and molecular docking study of *Horsfieldia iryaghedhi*

Rajapaksha HKK, Fernando MN, Nelumdeniya NRM, Bandara AWMKK, Silva ARN

## ABSTRACT

**Background:** Phytochemicals are still a valuable source to develop clinically important drugs in treating chronic and acute diseases. Inflammation is a response to an injurious stimulus of the body and novel therapeutic agents are needed to alleviate the condition with minimum side effects. **Aims and Objectives:** To investigate *in vitro* anti-inflammatory activity of methanol and aqueous leaf, bark, and combination extracts of plant *Horsfieldia iryaghedhi* by heat-induced egg albumin denaturation method and to analyze the phytochemicals of *Horsfieldia iryaghedhi* for their anti-inflammatory potential against cyclooxygenase- 2 (COX-2) using molecular docking. **Methodology:** Matured and fully expanded fresh leaves and barks of *H. iryaghedhi* were collected, and the extractions were obtained cold maceration using 99.9% methanol and distilled water as solvents. A concentration series was then developed, and the anti-inflammatory activity was evaluated against Diclofenac sodium as the positive control, using the heat-induced egg albumin denaturation method. Further, selected phytochemicals were tested against COX-2 enzyme (PDB ID: 5IKR) using site-specific molecular docking with autodock vina and the binding energies and pharmacokinetic and toxicity parameters were evaluated. **Results:** The methanol and aqueous extracts have shown a moderate to strong concentration-dependent anti-inflammatory activity with reference to standard Diclofenac sodium ($IC_{50}$ 116.4 μg/ml) and Methanol bark extract exhibited potent anti-inflammatory activity compared to other extracts ($IC_{50}$ 293 μg/ml). Further, Methanol and aqueous extracts showed a statistically significant correlation between concentration and percentage inhibition ($p<0.05$, $R^2 \approx 1$). The molecular docking results suggest that the phytochemicals available on the plant have possible COX-2 inhibitory activity and the compounds selected (Methyl 2,4-dihydroxy-6-methylbenzoate and N, N-Dimethyl-5-methoxy tryptamine) even got favourable toxicity and pharmacokinetic parameters confirming their drugability. **Conclusion:** Methanol bark extract of *Horsfieldia iryaghedhi* have marked *in vitro* anti-inflammatory activity. The results indicate a solid possibility of lead discovery of anti- inflammatory agents from the bark and leaves of *Horsfieldia iryaghedhi*. However, further molecular dynamics studies and *in vivo* tests may be required to confirm the findings.

## INTRODUCTION

The World Health Organization (WHO) estimates that 60% of people worldwide use herbal medicine for their basic medical requirements, and that value may increase by up to 80% in developing countries [1]. Even in developed countries, 10-50% of the population frequently consumes herbal products. Researchers suggest the primary rationale for using herbal medications in these countries is the expectation of improved tolerability compared to synthetic drugs. In underdeveloped nations, herbal medicines are sometimes the only available and affordable therapy option. So, using herbal products for therapeutic purposes is quite important worldwide [2].

Phytocompounds and their chemical analogs have produced many clinically helpful medications for treating chronic and acute disorders. According to the World Health Organization, trade in medicinal plants, raw herbal materials, and herbal pharmaceuticals is increasing by approximately 15% annually.

The growing popularity and acceptance of herbal therapy stem from the notion that all-natural products are safe, inexpensive, and readily available [3].

**Correspondence:**
**ARN Silva**
Department of Basic Sciences, Faculty of Allied Health Sciences, General Sir John Kotelawala Defence University, Weraera, Sri Lanka
Email: nsrajith@kdu.ac.lk, nsrajith2005@yahoo.com

The inflammatory process responds to an injurious stimulus that various infectious and physical agents can elicit. The host's inflammatory reaction is crucial for obtrusion and resolution of the infectious process. It is also responsible for the signs and symptoms of the disease condition and involves a complex series of host responses involving several biochemical pathways. Hence, the ineptitude inflammatory process may cause further damage to the tissues sometimes and has to be controlled [4].

The main drug classes used for the management of inflammatory diseases are steroids and non-steroidal anti-inflammatory drugs (NSAIDs). Even though NSAIDs have fewer side effects than steroids, they do cause inherent adverse effects, especially gastric ulcers [5]. Different NSAIDs can reduce pain and inflammation by blocking the metabolic process starting from arachnoid acid, especially by inhibiting the Cyclooxygenase enzymes (COXs); a major pathway that engages in the biosynthesis of thromboxane, prostaglandin, and prostacyclins [6]. Hence NSAIDs are used as an analgesic and anti-inflammatory agent in conditions such as pyrexia, gout, muscle pain, migraines, dysmenorrhea, arthritic conditions, and acute trauma.

Developing newer anti-inflammatory drugs with minor side effects is essential. It will enhance patient safety and compliance too. For this reason, in recent times, more interest has been shown in alternative and natural drugs for treating various diseases, but the scientific evidence is scarce [7].

*Horsfeildia iryaghedhi (H. iryaghedhi)*, often known as the Malaboda tree, is a flowering plant native to Sri Lanka and belongs to the family Myristicaceae. It is a fast-growing tree that reaches 10 to 20 meters in the wet zones, especially along the border of paddy fields, the margin of water streams and rivers. Various parts of the plant are widely prescribed for indigestion, dysentery, hiccough, wasting disease, and other medical conditions. Plants in the Myristicaceae family are well-known fragrant perennial plants with a distinctive aroma that has a variety of medicinal uses, including treating stomach ulcers, indigestion, and, liver diseases well as acting as an emmenagogue, nerve tonic, diuretic, diaphoretic and aphrodisiac. A famous plant from the same family; Nutmeg is most commonly used to prevent diarrhea and gastroenteritis. Additionally, it aids in the reduction of symptoms associated with digestive problems such as nausea, vomiting, and polyphagia [8]. Major members of the family Myristicaceae are; *Horsfieldia irya, Horsfieldia iryaghedhi, Myristica dactyloides,* and *Myristica fragrans. Horsfieldia irya* bark extraction by decoction is used to treat sore throat and latex is recommended to clean ulcers. Plant leaves can be used to abstract pus from boils and sores. Macerated root with lime juice is a remedy for toxic snake bites (9). Bark extraction of *Horsfieldia iryaghedhi* is widely prescribed for indigestion. Abdominal pain is most common in older adults. So, as a treatment seed ground of *H. iryaghedhi* with lime water is preferred. Dry flowers of male plants with bee honey are highly used in ayurvedic medicines to improve male sexual abilities. Flower and bark extracts are highly recommended in treating dysentery, hiccough, and wasting disease [9]. Bark and leaf extracts of *M. dactyloids* are preferred for throat ailments. Myristica fragrans is used to relieve flatulence and treat nausea and vomiting while the plant's nut is used to remedy loose bowels [9].

This plant may have many other medicinal uses that are yet to be found. Thus, this study aimed to discover the anti-inflammatory activity of *H. iryaghedhi* leaves and bark extract and a combination of them (leave and bark) using the egg albumin denaturation method as an indirect measure against inflammation.

Molecular docking is an important method structural molecular biology and computer-aided drug design. The purpose of molecular docking is to anticipate a ligand's preferred binding mode(s) with a protein with a known three-dimensional structure. This is a prescreening technique which help to discover lead molecules that can be used as future drugs. The ability of a protein (enzyme) and nucleic acid to interact with tiny molecules to form a supramolecular complex has a significant impact on protein dynamics, which can either increase or impede biological function. Molecular docking describes the behavior of tiny molecules in target protein binding sites [10].

## MATERIAL AND METHODS

**Materials**

*Chemicals*

Absolute Methanol (99% v/v) (Sigma Aldrich), Diclofenac Sodium (DC) (98% w/w) (standard) powder from the State Pharmaceutical Manufacturing Corporation of Sri Lanka. Gentamycin (80mg/2ml) (SPC), Phosphate buffered saline (pH 6.4), Normal saline (0.9% w/v) (B.BRAUN), Barium Chloride (1.175% w/v $BaCl_2.2H_2O$), Sulphuric Acid (1% v/v), Muller Hinton Agar (HiMedia Laboratories Pvt.Ltd).

*Instruments and Equipment*

Ultra-violet Spectrophotometer (Thermo scientific GENESYS), Rotary evaporator (HAHNSHIN Scientific-model no-H2005V,SR no;V-00449), Analytical balance (ACET, Model No: CY-224C, S/N: R600008446), Laboratory vortex mixture (Huma Twist, REF: 17175, S/N:VB192AH011635), Water bath (Equitron, S/N: NR11HD-18611), pH meter (EUTECH, S/N:02433), Autoclave machine (TOMYKOGYO co. ltd, Model-SX-500, SR no; 49133064), Hot air oven (BOV-V2225F with RS485), Incubator (CLW 240 IG SMART), Laboratory grinder( INNOVEX, Model: IMG O10).

**Collection and authentication of plant materials**

Fully expanded matured leaves and fresh bark of *Horsfieldia iryaghedhi* (RUK) were collected from the western province of Sri Lanka (Coordinates: 6°48'05.4"N, 79°58'37.6"E). The plant materials were properly dried and directed to the Pharmaceutical Botany division at Ayurveda Research Institute, Nawinna, Sri Lanka for taxonomical identification. A sample specimen has been preserved in the laboratory for future studies.

**Evaluation of *In vitro* anti-inflammatory activity**

*Preparation of crude extracts of plant materials*

Selected plant materials of *H. iryaghedhi* were thoroughly cleaned under running tap water and then with distilled water to remove any impurities. Then, the samples were air-dried until a constant weight was obtained. The dried plant materials were then powdered and extracted using water and methanol as extraction solvents.

Precisely weighted 20g of powdered plant materials of each leaf, bark, and combination (leaf + bark) were suspended in 160ml of 99.9% methanol in separate closed conical flasks. These were kept for seven days with occasional shaking in an orbital shaker. The extracts were then filtered through a double-layered muslin cloth and Whatman No 1 filter paper. Methanol was evaporated from the filtrate to get the dry residue using a rotary evaporator (HAHNSHIN Scientific-model no. H-2005V, SR no: V-00449). The samples were further dried at room temperature for 2 hours. The same procedure was carried out to prepare the aqueous extracts.

*Egg Albumin separation process for protein denaturation assay*

Egg whites were first separated from the yolk. 25 mL of egg white was measured and diluted with distilled water to obtain a 100 mL solution. It was mixed and stirred vigorously until notable quantities of whitish substance decreased. Then, the prepared egg white solution was centrifuged at 4000rpm for 20 minutes. The precipitated globulin was removed, and the resulting egg albumin solution was used for the albumin denature assay [11].

*Preparation of positive control/ standard (Diclofenac Sodium)*

The standard positive control, having 1000μg/mL (w/v) strength, was prepared by dissolving 0.25g of Diclofenac sodium in 50 mL of distilled water. From this stock solution, 4 mL was further diluted with 20 mL of distilled water to achieve a concentration of 1000μg/mL. This dilution process was extended to create a series of test solution concentrations at 1000μg/mL, 500μg/mL, 250μg/mL, 125μg/mL, 62.5μg/mL, 31.25μg/mL and 15.625μg/mL [12].

*Preparation of the negative control*

Negative control was prepared by combining 2.8 mL Phosphate Buffer Saline (PBS) (6.4 pH), 0.2 mL of egg albumin solution, and 2 mL of distilled water.

*Preparation of the serial dilutions*

Stock solutions with 1000μg/mL (w/v) of strength were prepared by dissolving 0.25g of each plant extract (methanolic leaf, bark, and combination leaf + bark, and bark, and aqueous leaf, bark, and combination leaf + bark) in 50 mL of distilled water. From these stock solutions, 4 mL was further diluted with 20 mL of distilled water to achieve a concentration of 1000μg/mL. This dilution process was extended to create a series of test solution concentration at 1000μg/mL, 500μg/mL, 250μg/mL, 125μg/mL, 62.5μg/mL, 31.25μg/mL, and 15.625μg/mL [13].

*Determination of the anti-inflammatory activity of Horsfeildia iryaghedhi using egg albumin denaturation assay*

The reaction mixtures were prepared using centrifuged egg albumin fraction (0.2mL), PBS of 6.4 pH (2.8mL) and 2mL of each different concentration of *H. iryaghedhi* aqueous extracts, and methanol extracts, and Diclofenac sodium separately. Then, the mixtures were incubated in a water bath at 37°C ± 2°C for 20 minutes. The temperature increased to 70°C, and the reaction mixtures were maintained for 5 minutes. Then the mixtures were cooled to room temperature, and absorption was measured using the UV spectrometer at 288 nm. The results were made in triplicate. Relevant intensities for the maximum wavelength were obtained to calculate the percentage inhibition.

The percentage inhibition of protein denaturation was calculated relative to the control using the following formula [14].

Percentage inhibition = $(V_t/V_c - 1) \times 100$

Where,

$V_t$ = Absorbance of the test sample

$V_c$ = Absorbance of control [14].

*Statistical analysis*

All data were analyzed statistically using SPSS version 23 and GraphPad Prism 9 (version 9.3.1). The descriptive data were expressed as mean ± standard error while the percentage of inhibition of different sample concentrations was analyzed by independent sample *t*-test. The differences were considered to be statistically significant when $P < 0.05$.

**In Silico screening**

*ADMET analysis*

ADMET analysis was performed using the SwissADME online server, and the toxicity data of each phytochemical was obtained from the Protox II online server. Lipinski's rule of five, ADME/T, and drug-like properties were focused on determining compounds' drugability.

*Molecular docking*

Molecular docking was performed for the selected phytochemicals in the plant *Horsfieldia iryaghedhi*. The 3D chemical structures of phytochemicals were obtained from the PubChem database and energy minimization was performed using Avogadro molecular visualization software using the steepest descent algorithm with MMFF94 forcefield. The crystallographic enzyme structure of the target human cyclooxygenase-2 enzyme (COX-2) was obtained from the rcsb PDB data bank (PDB ID 5IKR) and prepared for the docking using Auto Dock Tools (ADT) and Discovery Studio Visualizer. First, the incorporated ligand molecules and water were removed, and then the missing residues and atoms were introduced. Later, Discovery Studio Visualizer Softwears added to the protein structure, and Kollman charges were introduced. Site-specific Docking was performed for the selected phytochemicals using autodock vina [15]. The results and best docking poses were analyzed using Biovia Discovery Studio Visualizer.

**RESULTS AND DISCUSSION**

After many years of struggle against illnesses, people pursue drugs in seeds, bark, fruits, and other parts of the plants. The development of knowledge about the usage of medicinal plants has increased the ability of physicians and pharmacists to respond to emerging illnesses and spread professional services to facilitate human life [16]. Still, the traditional system of medicine is widely believed and followed by people worldwide [17].

Traditional remedies and medicinal plants are frequently used as a normative basis for maintaining good health in most underdeveloped countries. Medicinal plant usage to cure disease has been done since ancient times and it can be considered the origin of modern medicine [18]. Many medicinal plants are documented in Ayurveda and Unani medicine for treating various diseases. Scientists have identified significant amounts of medicinal properties of plants such as antimicrobial, anti-inflammatory, antioxidant, anti-cancer, and immunomodulatory which have become therapeutically crucial in curing diseases. Physicians involved in research have recently found that herbal plants can treat several types of chronic disorders such as asthma, diabetic mellitus, and cardiovascular diseases [19].

According to WHO data, around 21,000 plant species have the potential to be used as medicinal plants. The biggest advantage is that these remedies are in sync with nature. Being independent of any age group and sex when using herbal treatment can be considered a golden option [1].

Inflammation has direct etiological evidence in the pathology of certain diseases like rheumatoid arthritis, cancers, and even diabetes [20]. Since proteins are sensitive molecules, they denature by modifying the structures, making hydrogen, hydrophobic, and disulfide bonds. Inhibiting protein denaturation is a potential way to assess anti-inflammatory effectiveness [21]. Here, the anti-inflammatory activity of methanol and aqueous extracts of *H. iryaghedhi* was evaluated using an in vitro egg albumin denaturation assay, where anti-inflammatory activity is measured by the capacity to inhibit albumin denaturation [22]. If the plant extract contains anti-inflammatory phytoconstituents, it simply prevents protein denaturation. The activity of selected phytochemicals was further assessed *in silico* docking studies using COX-2 as the target and keeping mefenamic acid, ibuprofen, and diclofenac as reference drugs.

The protein denaturation was assessed using a UV-Vis Spectrophotometer in the 200nm to 800nm range. The optimum maximum wavelength of the denatured albumin protein was observed at 288nm. A UV-visible spectrum for egg albumin denaturation assay is shown in Figure 1. Considering the above fact, the study was further conducted from 200nm to 800nm, and the readings were acquired at their highest wavelength.

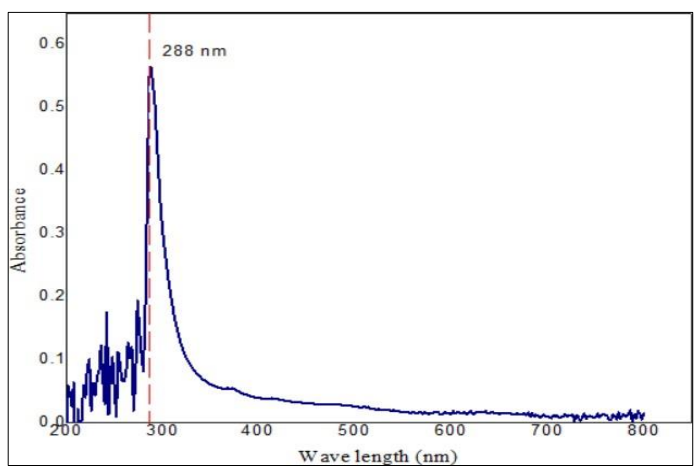

Figure 1: UV absorption spectrum of protein denaturation assay

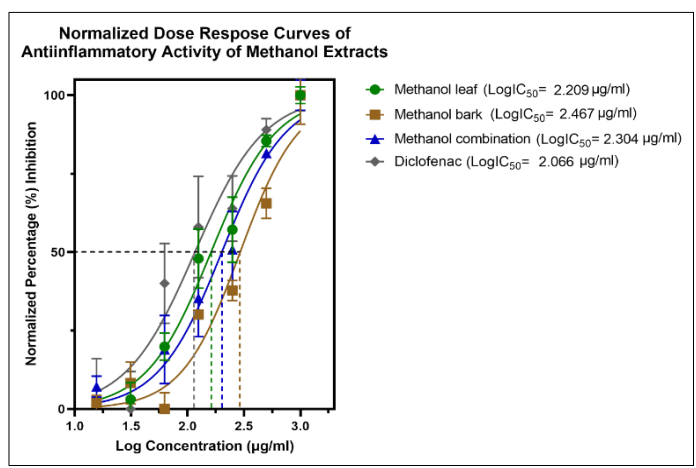

Figure 2: Dose-response curve for the anti-inflammatory activity of methanol extracts based on inhibition percentage

Dose-response data for methanol and aqueous leaf, bark, and combination extracts of *Horsfieldia iryaghedhi* and positive control are shown separately in Tables 1 and 2. The dose-response curve for methanol extracts and positive control are shown in Figure 2 while the dose-response curves for aqueous extract and positive control are shown in Figure 3. The methanol and aqueous extracts have shown a moderate to strong concentration-dependent anti-inflammatory activity concerning standard control (Diclofenac Sodium; $IC_{50}$ 116.4μg/ml). Interestingly Methanol bark extract exhibited potent anti-inflammatory activity compared to other extracts ($IC_{50}$ 293.0μg/ml). Methanol and aqueous extracts showed a statistically significant correlation between concentration and percentage inhibition ($p<0.05$, $R^2 \approx 1$).

**Table 1:** Mean percentage inhibition of egg albumin denaturation with positive control, methanol leaf, methanol bark, and methanol leaf/bark combination extract of *H. iryaghedhi* at different concentrations

| Concentration (μg/mL) | Mean% inhibition ± SEM | | | |
|---|---|---|---|---|
| | Positive control | Methanol leaf | Methanol Bark | Methanol (Leaf/bark) combination |
| 15.625 | 34.8±9.67 | 14.4±1.56 | 36.9±1.49 | 17.5±2.30 |
| 31.250 | 33.8±8.16 | 16.3±3.25 | 40.7±4.09 | 12.6±2.91 |
| 62.500 | 54.3±8.68 | 26.7±2.66 | 35.7±3.18 | 25.6±7.48 |
| 125.000 | 63.5±11.0 | 44.0±5.81 | 54.1±1.98 | 36.9±8.46 |
| 250.000 | 66.5±7.12 | 49.7±6.39 | 58.8±2.96 | 47.7±8.38 |
| 500.000 | 79.4±2.41 | 67.1±1.14 | 75.8±2.96 | 68.7±1.02 |
| 1000.000 | 85.0±3.37 | 79.1±1.64 | 97.0±5.67 | 81.4±3.30 |

According to Table 1, with methanol leaf, bark, and combination extracts of *H. iryaghedhi*, the percentage inhibition of egg albumin denaturation ranged from 14.4±1.56% to 79.1±1.64%, 36.9±1.49% to 97.0±5.67% and 17.5±2.30% to 81.4±3.30% respectively at the concentration series of 15.625 to 1000μg/mL.

A positive correlation was identified between the percentage of protein inhibition and concentrations of the plant extracts. The bark of *H. iryaghedhi* is rich with compounds with anti-inflammatory activity, unlike leaves. It was confirmed when comparing the percentage inhibition of protein denaturation data. For example, 1000μg/mL methanol bark extract of *H. iryaghedhi* exhibited the highest percentage inhibition of 97.0±5.67%, while leaf extract of the same concentration showed only a percentage inhibition of 79.1±1.64%. Combining bark with the leaves increases the value up to 81.4±3.30%. The results showed that the anti-inflammatory activity of *H. iryaghedhi* methanol extracts were concentration-dependent, and $IC_{50}$ values of methanol leaf, methanol bark, and methanol combination were obtained as 161.7μg/mL, 293.0μg/mL, and 201.4μg/mL respectively.

It revealed that the methanol extracts of *H. iryaghedhi* showed a solid ability to protect albumin from denaturation, which translates to intense anti-inflammatory activity. The $r^2$ values of methanol leaf, bark, and combination extracts were 0.9263 ($p<0.05$), 0.9087 ($p<0.05$), and, 0.8792 ($p<0.05$), respectively. The $r^2$ value of the reference drug was 0.7913 ($p<0.05$), showing that the observed concentration-dependent is confirmed as authentic and statistically significant.

**Table 2:** Mean percentage inhibition of egg albumin denaturation with positive control, aqueous leaf, aqueous bark, and aqueous (leaf/bark) combination extract of *H. iryaghedhi* at different concentrations

| Concentration (μg/mL) | Mean % inhibition | | | |
|---|---|---|---|---|
| | Positive control | Aqueous leaf | Aqueous bark | Aqueous (leaf/bark) combination |
| 15.625 | 34.7±9.67 | 34.9±2.73 | 39.2±3.99 | 41.4±4.56 |
| 31.250 | 33.8±8.16 | 52.0±3.86 | 30.0±8.31 | 34.2±1.83 |
| 62.500 | 54.3±8.68 | 43.0±1.64 | 28.8±3.91 | 41.1±9.68 |
| 125.000 | 63.5±11.0 | 41.8±5.36 | 43.5±1.92 | 41.9±1.69 |
| 250.000 | 66.5±7.12 | 52.4±0.97 | 54.8±5.30 | 54.2±1.36 |
| 500.000 | 79.4±2.41 | 69.3±3.78 | 47.8±4.51 | 54.8±1.24 |
| 1000.000 | 85.0±3.37 | 80.4±2.55 | 67.6±5.53 | 77.0±2.49 |

According to Table 2, with aqueous leaf, bark, and combination extract of *H. iryaghedhi*, the percentage inhibition of egg albumin denaturation ranged from 34.9±2.73% to 80.4±2.55%, 39.2±3.99% to 67.6±5.53% and 41.5±4.56% to 77.0±2.49% respectively at concentration series of 15.625 to 1000μg/mL.

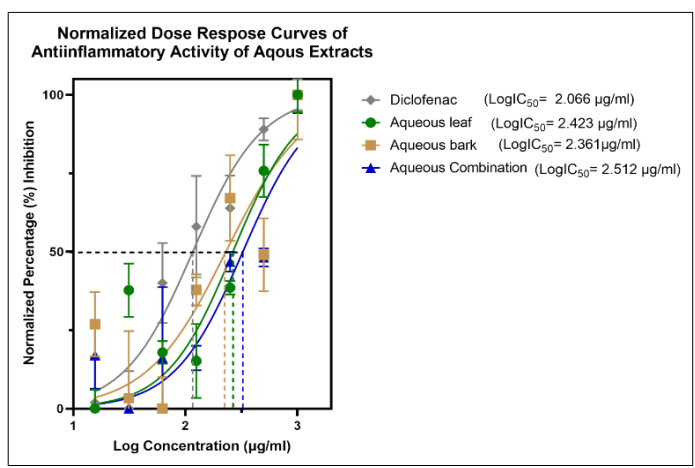

**Figure 3:** Dose-response curve for the anti-inflammatory activity of aqueous extracts based on inhibition percentage

Aqueous leaf extract of *H. iryaghedhi* exhibited the highest percentage inhibition of 80.4±2.55% in its highest concentration (1000µg/mL). Surprisingly, bark extract exhibited only 67.6±5.53% in the same concentration. However, the aqueous combination exhibited 77.0±2.49% inhibition. It can be concluded that the aqueous leaf extract has shown the highest percentage of inhibition compared to extracts of aqueous bark and combination. As the results showed, the anti-inflammatory activity of its *H. iryaghedhi* aqueous extracts was concentration- dependent and $IC_{50}$ values of aqueous leaf, aqueous bark, and aqueous combination were obtained as 265.1µg/mL, 229.7µg/mL and 325.3µg/mL respectively.

It shows that the aqueous extracts of *H. iryaghedhi* showed a moderate to strong ability to protect albumin from denaturation, which translates to moderate to strong anti-inflammatory activity. The $r^2$ values of aqueous extracts of *H. iryaghedhi* were 0.7277 ($p<0.05$), 0.5729 ($p<0.05$), and, 0.7148 ($p<0.05$), respectively. The $r^2$ value of the positive control was 0.7913 ($p<0.05$), showing that the observed results are concentration-dependent and statistically significant.

With this experimental finding it was confirmed that, the methanol extract of *Horsfieldia iryaghedhi* exhibited a significant percentage inhibition of protein denaturation compared to aqueous extracts. Methanol bark extract presented the highest ability to inhibit albumin denaturation and it is greater than the percentage inhibition of Diclofenac sodium (97.0±5.67 vs 85.0±3.37%) in its highest concentration (1000µg/mL). Methanol bark extract showed a higher potency than Diclofenac sodium.

The Myristicaceae family comprises more than 80 species of the genus *Horsfieldia,* and most of the plants are widely used for their antimicrobial and anti-inflammatory properties in cosmetics and dermatological treatments [23]. The healing properties of medicinal plants are primarily due to numerous secondary metabolites such as lignans, flavons, sterols, alkaloids, and essential oils [24]. Phytochemical screening of *Horsfieldia iryaghedhi* leaf, bark, and timber has confirmed the occurrence of d-asarinin, dihydrocubebin, dodycanoylphloroglucinol, sitosterol, myristic acid, and trimyristin [25] and authors suggest that these compounds may lead to significant positive anti-inflammatory activity. As the next step, the most abundant phytochemicals in *H. iryaghedhi* were subjected to Molecular docking using Autodock vina.

Molecular docking is a Structure-Based Drug development method that reveals bind- ing site topology (presence of clefts, cavities, sub-pockets, and electrostatic properties, such as charge distribution) which helps to prescreen molecules as drug candidates [26]. The binding affinity of selected phytochemicals were tested against cyclooxygenase-2 enzyme (COX-2) (PDB ID 5IKR) and the results were compared with three conventional drugs Mefenamic acid, Diclofenac and Ibuprofen. Further, the drug likeliness and toxicity of selected phytochemicals were also assessed.

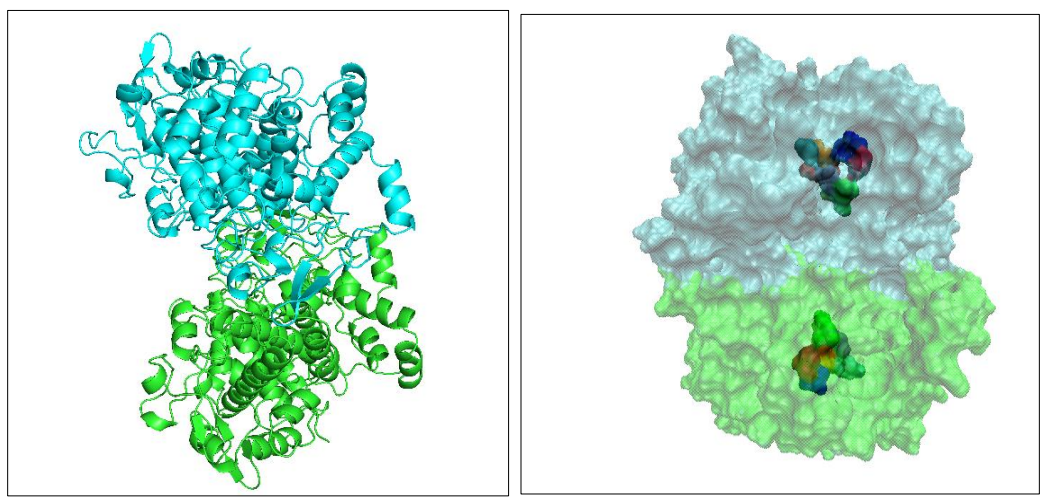

A.           B.
**Figure 4: a)** Human cyclogenase II enzyme (PDB-5IKR) b) Represent at ions of the active sites

**Table 3:** ADMET Analysis data

| Phytochemical Name | Lipinski's rules | | | | Lipinski's Rule violations | GI absorption | BBB permeability | CYP3A4 inhibitor | Bioavailability Score |
|---|---|---|---|---|---|---|---|---|---|
| | MW < 500 | HBA ≤ 10 | HBD ≤ 5 | LogP ≤ 5 | | | | | |
| Myristic acid | 228.37 | 2 | 1 | 3.69 | 0 | High | Yes | No | 0.85 |
| Trimyristin | 723.16 | 6 | 0 | 7.88 | 2 | Low | No | No | 0.17 |
| Asarinin | 354.35 | 6 | 0 | 1.98 | 0 | High | Yes | Yes | 0.55 |
| Horsfieldin | 356.37 | 6 | 1 | 1.57 | 0 | High | Yes | Yes | 0.55 |
| Phloroglucinol | 126.11 | 3 | 3 | 0.18 | 0 | High | Yes | Yes | 0.55 |
| Dihydrocubebin | 358.39 | 6 | 2 | 1.98 | 0 | High | No | Yes | 0.55 |
| Sitosterol | 414.71 | 1 | 1 | 6.73 | 1 | Low | No | No | 0.55 |
| (S)-3-Hexyl-5,6-dihydro-6-undecyl-2H-pyran-2-one | 336.55 | 2 | 0 | 5.08 | 1 | Low | No | No | 0.55 |
| Viridiflorol | 222.37 | 1 | 1 | 3.81 | 0 | High | Yes | No | 0.55 |
| Hexadecanoic acid | 396.69 | 2 | 1 | 6.41 | 1 | Low | No | No | 0.85 |
| Methyl 2,4-dihydroxy-6-methyl benzoate | 182.17 | 4 | 2 | 1.06 | 0 | High | Yes | No | 0.55 |
| Methyl 2,4-dihydroxy-3,6-dimethyl benzoate | 254.24 | 4 | 2 | 1.08 | 0 | High | Yes | Yes | 0.55 |
| 4',7-Dihydroxyflavone | 272.3 | 4 | 2 | 1.87 | 0 | High | Yes | Yes | 0.55 |
| Catechin | 290.27 | 6 | 5 | 0.24 | 0 | High | No | No | 0.55 |
| Epicatechin | 290.27 | 6 | 5 | 0.24 | 0 | High | No | No | 0.55 |
| 2,4'-Dihydroxy-4-methoxydihydrochalcone | 272.3 | 4 | 2 | 1.91 | 0 | High | Yes | Yes | 0.55 |
| 7,3',4'-Trihydroxyflavone | 270.24 | 5 | 3 | 0.52 | 0 | High | No | Yes | 0.55 |
| 4'-Hydroxy-7-methoxyflavone | 268.26 | 4 | 1 | 1.33 | 0 | High | Yes | Yes | 0.55 |
| Palmitic Acid | 256.42 | 2 | 1 | 4.19 | 1 | High | Yes | No | 0.85 |
| 3,4-Dihydroxybenzoic acid | 154.12 | 4 | 3 | 0.4 | 0 | High | No | Yes | 0.56 |
| Horsfiline | 232.28 | 3 | 1 | 1.14 | 0 | High | No | No | 0.55 |
| 1,2,3,4-Tetrahydro-2-methyl-6-methoxy-beta-carboline | 202.25 | 2 | 2 | 1.22 | 0 | High | Yes | No | 0.55 |
| N, N-Dimethyl-5-methoxy tryptamine | 218.29 | 2 | 1 | 1.5 | 0 | High | Yes | No | 0.55 |
| Cardinal | 222.37 | 1 | 1 | 3.67 | 0 | High | Yes | No | 0.55 |
| Germacrene D | 204.35 | 0 | 0 | 4.53 | 1 | Low | No | No | 0.55 |
| humulene | 204.35 | 0 | 0 | 4.53 | 1 | Low | No | No | 0.55 |
| 1-2-6-Dihydroxyphenyldodecan-1-one | 292.42 | - | - | - | - | - | - | - | - |
| 6-Methoxy-2-methyl-1-3-4-9-tetrahydropyran-3-4-indole | 202.25 | - | - | - | - | - | - | - | - |
| Dodecanoylphloroglucinol | 230.26 | - | - | - | - | - | - | - | - |

**Table 4:** Toxicity prediction data of the selected compounds as per Protox server results

| Phytochemical Name | Predicted LD$_{50}$ (mg/kg) | Predicted toxicity class | Average similarity | Prediction accuracy | Hepatotoxicity | Carcinogenicity | Immunotoxicity | Mutagenicity | Cytotoxicity |
|---|---|---|---|---|---|---|---|---|---|
| Myristic acid | 900 | 4 | 100% | 100% | - | - | - | - | - |
| Trimyristin | - | - | - | - | - | - | - | - | - |
| Asarinin | 1500 | 3 | 72.19% | 69.26% | - | + | + | - | - |
| Phloroglucinol | 200 | 3 | 100% | 100% | - | - | - | - | - |
| Dihydrocubebin | 720 | 4 | 80.37% | 70.97% | - | - | - | - | - |
| Sitosterol | 890 | 4 | 89.38% | 70.97% | - | - | - | - | - |
| (S)-3-Hexyl-5,6-dihydro-6-undecyl-2H-pyran-2-one | 1890 | 4 | 81.49% | 70.97% | - | - | - | - | - |
| Viridiflorol | 2000 | 4 | 81.25% | 70.97% | - | - | - | - | - |
| Hexadecanoic acid | 90 | 4 | 100% | 100% | - | - | - | - | - |
| Methyl 2,4-dihydroxy-6-methyl benzoate | 887 | 4 | 77.26% | 69.26% | - | - | - | - | - |
| Methyl 2,4-dihydroxy-3,6-dimethyl benzoate | 1900 | 4 | 77.35% | 69.26% | - | - | - | - | - |
| 4',7-Dihydroxyflavone | 2500 | 5 | 83.2% | 70.97% | - | - | - | - | - |
| Catechin | 10000 | 6 | 100% | 100% | - | - | - | - | - |
| Epicatechin | 10000 | 6 | 100% | 100% | - | - | - | - | - |
| 2,4'-Dihydroxy-4-methoxydihydrochalcone | 500 | 4 | 76.13% | 69.26% | - | - | - | - | - |
| 7,3',4'-Trihydroxyflavone | 1070 | 4 | 80.45% | 70.97% | - | - | - | - | - |
| 4'-Hydroxy-7-methoxy flavone | 4000 | 5 | 96.51% | 72.9% | - | - | - | - | - |
| Palmitic Acid | 900 | 4 | 100% | 100% | - | - | - | - | - |
| 3,4-Dihydroxybenzoic acid | 2000 | 4 | 87.23% | 70.97% | - | + | - | - | - |
| Horsfiline | 2 | 1 | 63.31% | 68.07% | - | - | - | - | - |
| 1,2,3,4-Tetrahydro-2-methyl-6-methoxy-beta-carboline | - | - | - | - | - | - | - | - | - |
| N,N-Dimethyl-5-methoxy tryptamine | 963 | 4 | 66.03% | 68.07% | - | - | + | - | - |
| Cadinol | 2830 | 5 | 96.55% | 72.9% | - | - | + | - | - |
| Germacrene D | 5300 | 5 | 80.77% | 70.97% | - | - | + | - | - |
| humulene | 3650 | 5 | 86.36% | 70.97% | - | - | - | - | - |
| 1-2-6-Dihydroxyphenyldodecan-1-one | - | - | - | - | - | - | - | - | - |
| 6-Methoxy-2-methyl-1-3-4-9-tetrahydropyrido-3-4-indole | - | - | - | - | - | - | - | - | - |
| Dodecanoylphloroglucinol | - | - | - | - | - | - | - | - | - |

**Table 5:** Binding interactions of the selected compounds with human cyclooxygenase-2 (PDB ID 5IKR) enzyme

| Phytochemical Name | Binding Energy (Kcal mol$^{-1}$) | Hydrogen bond interactions, | | Hydrophobic interactions |
|---|---|---|---|---|
| | | Classical | Non-classical | |
| Myristic acid | -5.095 | - | HIS207 | ILE408, VAL295, HIS388, LEU391, VAL447, LEU294 |
| Trimyristin | - | - | - | - |
| Asarinin | -8.697 | SER581 | GLN192, GLY354, HIS90 | |
| Horsfieldin | -8.175 | PHE580, ASN101, HIS356, ASN104, | HIS351, ASP347 | GLN350, |
| Phloroglucinol | - | - | - | - |
| Dihydrocubebin | -7.764 | HIS386, ASN382 | TRP387 | HIS388, |
| Sitosterol | -8.742 | SER119 | | TRP100, VAL89, TYR115, VAL116, ILE92, PHE96, PHE99 |
| (S)-3-Hexyl-5,6-dihydro-6-undecyl-2H-pyran-2-one | - | - | - | - |
| Viridiflorol | -6.393 | - | - | PRO86, VAL89 |
| Hexadecanoic acid | -5.617 | HIS214, THR212 | - | LEU391, HIS388, ALA202, PHE210, HIS207, **TYR385**, HIS386 |
| Methyl 2,4-dihydroxymethyl benzoateate | -6.463 | **SER530** | VAL349 | LEU352, ALA527 |
| Methyl 2,4-dihydroxy-3,6-dimethylbenzoate | -6.84 | | GLN203 | ALA202, HIS388 |
| 4',7-Dihydroxyflavone | - | - | - | - |
| Catechin | -8.049 | ASN382, THR206, GLN203, ALA199, | - | HIS386, HIS207 |
| Epicatechin | - | - | - | - |
| 2,4'-Dihydroxy-4-methoxydihydrochalcone | - | - | - | - |
| 7,3',4'-Trihydroxyflavone | - | - | - | - |
| 4'-Hydroxy-7-methoxy flavone | -7.911 | - | - | LEU352, GLY526, LEU359, ALA527, VAL349, LEU531, ILE345, VAL523, MET113 |
| Palmitic Acid | - | - | - | - |
| 3,4-Dihydroxybenzoic acid | -6.201 | ALA202 | GLN203 | |

| Horsfiline | -7.254 | THR206 | | HIS207, HIS388, HIS386 |
| --- | --- | --- | --- | --- |
| 1,2,3,4-Tetrahydro-2-methyl-6-methoxy-beta-carboline | - | - | - | - |
| N, N-Dimethyl-5-methoxy tryptamine | -6.852 | **TYR385**, MET522 | | VAL344, VAL523, ALA527, TYR348, LEU352, VAL349 |
| Cadinol | -6.876 | GLU524 | | LYS83, PRO84, LEU93, VAL89, VAL116 |
| Germacrene D | -6.541 | - | | TYR115, ILE112, TRP100 |
| Humulene | -6.556 | - | | VAL89 |
| 1-2-6-Dihydroxyphenyldodecan-1-one | -6.277 | - | ALA527, **SER530** | LEU531, VAL116, LEU359, VAL349, LEU352, **TYR385**, TRP387, PHE518, VAL523, MET522 |
| 6-Methoxy-2-methyl-1-3-4-9-tetrahydropyran-3-4-indole | -7.025 | - | | VAL349, GLY526, ALA527, LEU352, LEU384, TRP387, MET522 |
| Dodecanoylphloroglucinol | -6.788 | - | ALA527 | LEU531, VAL349, PHE209, PHE381, LEU534, VAL344, PHE205, TYR385, TYR348, LEU352 |
| Diclofenac | -7.533 | ARG120, TYR355 | | TRP387, LEU531, ALA527, VAL349, LEU352 |
| Mefenamic acid | -9.378 | **TYR385, SER530** | | LEU352, VAL349, ALA527, LEU531, MET522 |
| Ibuprofen | -7.52 | **TYR385** | **SER530** | LEU359, LEU531, VAL349, VAL116, ARG120, VAL523, GLY526, ALA527, LEU352 |

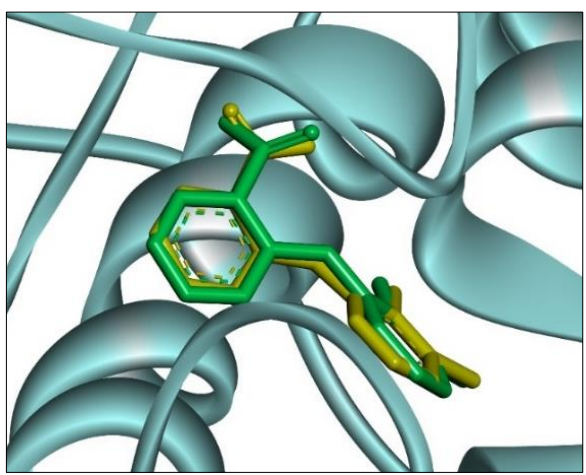
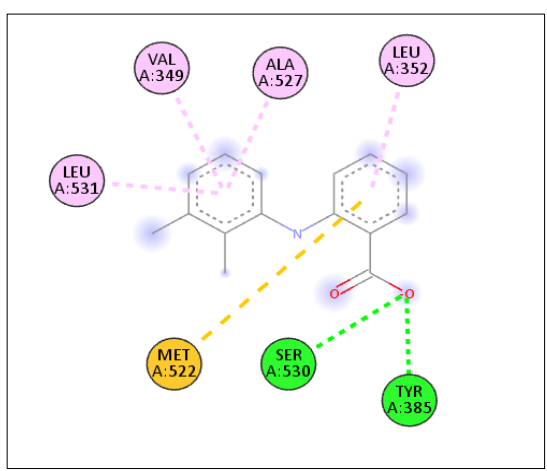

**Figure 5 - (a)**                                                                                      **Figure 5 - (b)**

**Figure 5 – a)** Mefanamic acid in the active site of COX 2 (PDB ID: IKR); Yellow- Mefanamic acid in the crystal structure, Green- Mefanamic acid in its best-docked pose. b) A schematic two-dimensional (2D) display of mefenamic acid, *Vanderwaal interactions are shown in purple in color; hydrogen bonds are green in color and Pi bonds are orange in color.

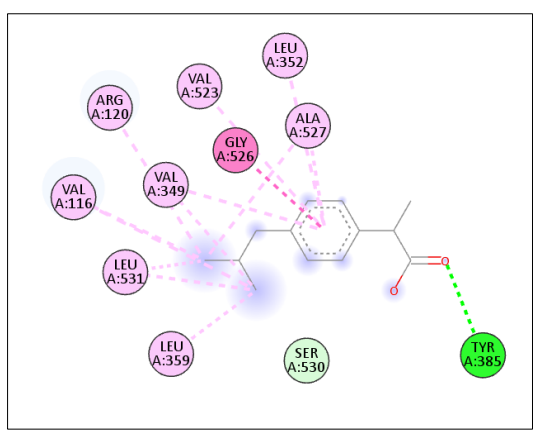
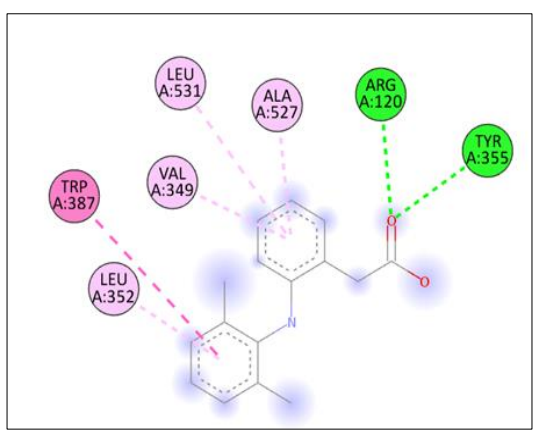

**a. Ibuprofen**                                                                                      **b. Diclofenac**

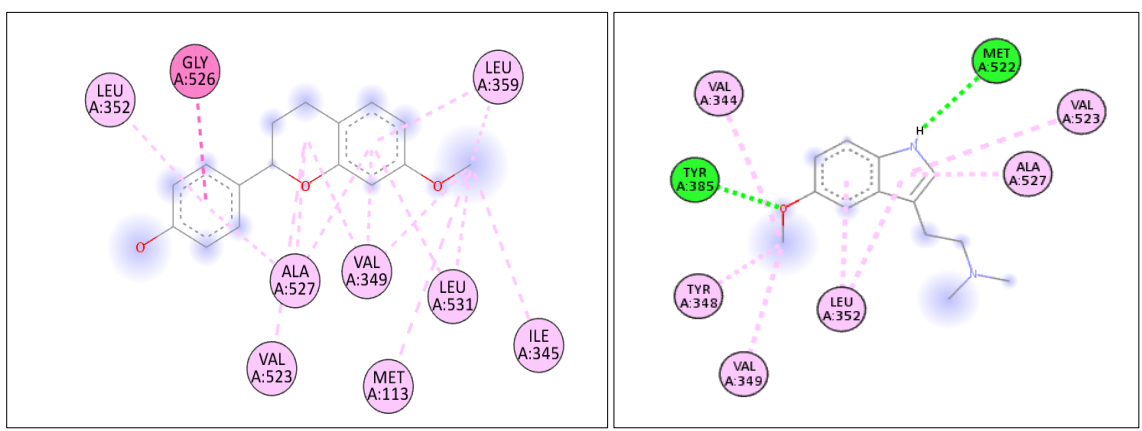

**c. 4-hydroxy-7-methoxyflurane N,**     **d. N-Dimethyl-5-methoxy tryptamine**

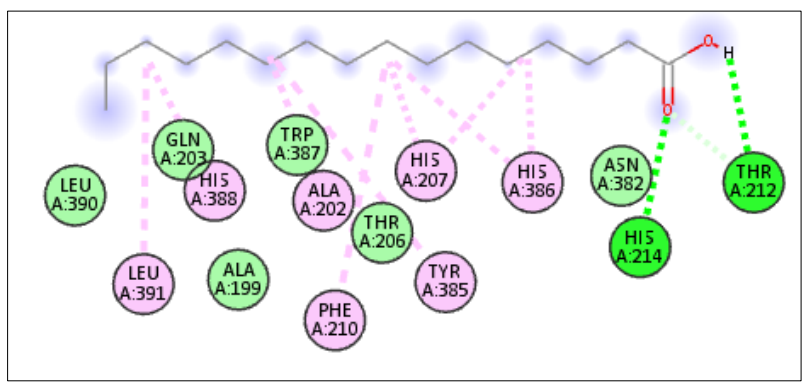

**d. Hexadecanoic acid**

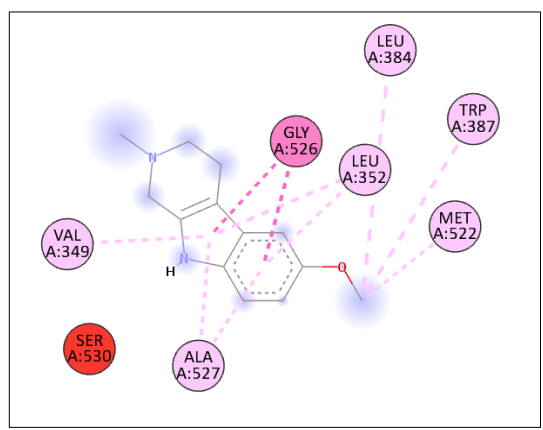

**e. Methyl 2,4-dihydroxy-6-methyl benzoate**

**Figure 6:** A schematic two-dimensional (2D) display of a) Ibuprofen b) Diclofenac c) 4-hydroxy-7-methoxyflurane d) N, N-Dimethyl-5-methoxy tryptamine e) Hexadecanoic acid; f) Methyl 2,4-dihydroxy-6-methyl benzoate, *Vanderwaal interactions are shown in purple in color; hydrogen bonds are in green in color and Pi bonds are in orange in colour.

The stability of the best-docked pose is determined by bonds created between the ligand and the critical amino acids involved. The hydrogen bond interaction is significant for the bioactivity followed by hydrophobic interactions. For determining the optimal ligand binding conformation, the pose with the lowest binding energy was taken into consideration. Other non-bonded interactions, such as hydrophobic bonding, were found in addition to hydrogen bonding interactions. All the docked compounds showed low binding energies in the range of -5.095- -9.378 Kjmol-1 with the active site of COX- 2 enzymes. NSAIDs that are commercially available, such as diclofenac, ibuprofen, and mefenamic acid, were utilized as benchmarks. Mefanamic acid demonstrated the best docking score of all the compounds, exhibiting a virtually exact docking pose to the crystal structure, indicating the accuracy of the docking method.

Moreover, it forms hydrogen bonds with the binding site's essential amino acids, TYR385 and SER 530, and mediates the COX-2 inactivation [27]. Diclofenac interacts with TYR 385 and Ibuprofen with other amino acids in the binding area, including TYR 355 and ARG120, to form hydrogen bonds. Furthermore, non-bonded interactions that these three molecules have with TRP387, LEU531, ALA527, VAL349, LEU352, MET52, LEU359, VAL116, ARG120, VAL523, and GLY526 are also present. Methyl 2,4-dihydroxy-6-methylbenzoate (-6.463 Kcal mol$^{-1}$), N, N-Dimethyl-5-methoxy tryptamine (-6.852 Kcal mol$^{-1}$), Hexadecanoic acid (-5.617 Kcal mol$^{-1}$) interact with the binding pocket amino acids including the crucial one which is TYR385**.** Methyl 2,4-dihydroxy-6-methylbenzoate and N, N-Dimethyl-5-methoxy tryptamine had zero violations of Lipinski's Rule and high GI absorption and reasonable bioavailability scores.

Although the majority of the phytochemicals exhibit binding to the selected active site of COX2, their huge structures with sidechains prevent them from interacting as predicted with the designated binding pocket. Nevertheless, the inhibition cannot be verified since they do not interact with the amino acids where the reference molecules interact. Phytochemicals such as δ-Cadinene, 1-2-6-Dihydroxyphenyldodecan-1-one, 6-Methoxy-2-methyl-1-3-4-9 tetrahydropyran-3-4_indole, dodecanoylphloroglucinol, Hexadecanoic_acid, and 4'-Hydroxy-7-methoxy flavone showed reasonable docking scores even though they only form non-bonded interactions with the other amino acids available in the binding pocket excluding TYR 385 and SER 530. Further, they have acceptable toxicity and pharmacokinetic data too. Asarinin, Dihydrocubebin, 4'-Hydroxy-7-methoxy flavone, Horsfieldin, and beta-Sitosterol showed greater predicted binding energies than the control drug diclofenac.

## CONCLUSION

This study showed that methanol and aqueous extract of bark, leaf, and a combination of *H. iryaghedhi,* an endemic plant in Sri Lanka, have exhibited marked concentration-dependent *in vitro* anti-inflammatory activity in the egg albumin denaturation assay. The bark showed higher activity compared to the other extracts. The molecular docking suggests that the phytochemicals (Methyl 2,4-dihydroxy-6-methylbenzoateand N,N-Dimethyl-5-methoxy tryptamin) available on the plant have possible COX-2 inhibitory activity and the compounds selected even got favorable toxicity and pharmacokinetic parameters confirming their drugability. The results indicate a solid possibility of lead discovery of anti-inflammatory agents from the bark and leaves of *Horsfieldia iryaghedhi*. However, further molecular dynamics studies and *in vivo* tests may be required to confirm the findings.

**Acknowledgments**

The authors would like to thank all the laboratory staff members of the Department of Pharmacy and Department of Basic Sciences, General Sir John Kotelawala Defence University, Werahera, Sri Lanka.

**Conflict of interest**

The authors declared no conflict of interest.

**Financial Support**

None declared.

**ORCID ID**

Hirunika Kavindi Kaushalya Rajapaksha: https://orcid.org/0009-0004-2711-5278

Maleesha Nathaliya Fernando: https://orcid.org/0009-0006-1234-5956

Rumesh Madhushanka Nelumdeniya: https://orcid.org/0000-0002-2788-207X

Kanchana Kumari Bandara: https://orcid.org/0000-0003-0371-6760

Rajith Neloshan Silva: https://orcid.org/0000-0001-9612-0733